\documentclass{emulateapj}
\usepackage{psfig}

\newcommand{\be}{\begin{equation}}
\newcommand{\ba}{\begin{eqnarray}}
\newcommand{\ee}{\end{equation}}
\newcommand{\ea}{\end{eqnarray}}  
\newcommand{\etal}{et al.\ }
\def\gtsima{$\; \buildrel > \over \sim \;$}
\def\ltsima{$\; \buildrel < \over \sim \;$}
\def\gsim{\lower.5ex\hbox{\gtsima}}
\def\lsim{\lower.5ex\hbox{\ltsima}}
\def\simgt{\lower.5ex\hbox{\gtsima}}
\def\simlt{\lower.5ex\hbox{\ltsima}}
\def\simpr{\lower.5ex\hbox{\prosima}}

\def\msun{\,{\rm M_\odot}}

\def\eg{{\frenchspacing\it e.g., }}
\def\E3{{\cal E}_{\rm g}^{III}}

\begin{document}
\submitted{}
\title{The Type Ia Supernova Rate}
\author{Evan Scannapieco \& Lars Bildsten}
\affil{ Kavli Institute for Theoretical Physics,
Kohn Hall, University of California,
Santa Barbara, CA 93106; \newline evan@kitp.ucsb.edu, 
bildsten@kitp.ucsb.edu}

\begin{abstract}

We explore the idea that the Type Ia supernovae (SN Ia) rate consists
of two components: a prompt piece that is  proportional to the star
formation rate (SFR) and an extended piece that is  proportional to the total
stellar mass.  We fit the parameters of this model to the local
observations by Mannucci  and collaborators and then study its impact
on  three important problems.   On cosmic scales, the model reproduces
the observed SN Ia rate density below $z=1$, and predicts that it
will track the measured SFR density at higher redshift, reaching a
value of 1-$3.5 \times10^{-4}$ yr$^{-1}$ Mpc$^{-3}$ at $z=2$.  In
galaxy clusters, a large prompt contribution helps explain the Fe
content of the intracluster medium.  Within the Galaxy, the model
reproduces the observed stellar [O/Fe] abundance ratios if we allow a
short ($\approx 0.7$ Gyr) delay in the prompt component.  Ongoing
medium-redshift  SN surveys will yield more accurate parameters 
for our model.

\end{abstract}


\section{Introduction}

Type Ia supernovae (SNe Ia) play a pivotal role in astrophysics.  On
cosmological scales they serve  as unparalleled distance indicators,
providing direct evidence that the low-redshift universe is
accelerating (Riess \etal 1998; Perlmutter \etal 1999). On galactic
scales, they act as the primary source of iron, producing $\approx 0.7
\msun$ per event (Tsujimoto \etal 1995), roughly an order of
magnitude more than in core-collapse SNe (Hamuy 2005). On stellar
scales, they represent an excellent example of explosive nuclear
burning, resulting in radioactively-powered light curves (\eg Nomoto
\etal 1984; Woosley 1990).

Nevertheless, many mysteries remain. While 
the peak magnitude and the decay time of SN Ia light curves
are tightly correlated
(Pskovskii 1977; Phillips 1993; Hamuy \etal 1995), the origin of
this relation is poorly understood (see discussion in Pinto \&
Eastman 2000; Mazzali et al  2001).  Similarly, while there is a
consensus that SNe Ia originate from thermonuclear ignition and 
burning of a C/O white dwarf in a binary system, it is
uncertain whether they are triggered by  accretion from a hydrogen-rich
companion  or from a merger with another white dwarf (see Branch \etal
1995).  This uncertainty in the progenitor makes it difficult to predict the
Ia rate in galaxies of varying masses,
ages, and star formation rates. 

Consequently, a wide range of models of this rate have been developed
(\eg Matteucci \& Recchi 2001; Greggio 2005). 
These are commonly parameterized by a delay
function, whose convolution with the star formation rate (SFR) yields
 the SN Ia rate. In principle, this is a completely general approach, as
the SN Ia rate must depend on the mass and age of the underlying stars.  
In practice, most of this generality is lost to the
assumption of a single  ``delay time'' (\eg Madau \etal 1998; Dahl\' en
\& Fransson 1999;  Gal-Yam \& Maoz 2004; Strolger \etal 2004).  These
fits are used to draw conclusions about SN Ia progenitors,
the most recent example of which is the claim that there must be a 2-4
Gyr delay in all SNe Ia relative to the burst of star formation
(Strolger et al. 2004).

However, this approach neglects the possibility that multiple
evolutionary  paths lead to SNe Ia. Indeed, there is direct
observational evidence that this is the case. The brightest events
(such as 1991T) only  occur in actively star-forming galaxies,  while
substantially under-luminous events (such as 1991bg)  are most
prevalent in E/S0 galaxies  (Hamuy \etal 1996; Howell 2001; van den
Bergh \etal 2005).  This is  an important clue that SNe Ia have at least
two evolutionary channels with different characteristic times:
one ``prompt,'' basically tracking the current SFR, and another so
delayed that it simply scales with the stellar mass (much as is
seen in accreting binaries with low-mass companions, such as
Cataclysmic Variables [Townsley and Bildsten 2005] or low-mass X-ray
binaries in E/S0s [Gilfanov 2004]).  We show here, that in addition to
explaining the SN Ia rates seen in nearby galaxies as described
by Mannucci \etal (2005, hereafter M05), such a simple two-component
model also resolves a wide range of outstanding issues.

We begin in \S2 by presenting the model, fitting the constants to
observations of SNe Ia in nearby galaxies, and stating a few of the
immediate repercussions.  In \S3 we study the implications of this
model in three important contexts: the iron content of galaxy
clusters,   the evolution of the average cosmic SN Ia rate density,
and the [O/Fe] abundance ratios of Galactic stars.  In \S4 we contrast
this approach with other models, and  we conclude  in \S 5.

\section{The Two-Component Model}

Following M05, we assume that there are two avenues for SNe Ia. 
Specifically, we adopt a model in which the 
SN Ia rate is the sum of two components: a
term proportional to the total stellar mass, $M_\star(t)$ 
(regardless of its age) and a term proportional  to the instantaneous 
SFR, $\dot M_\star(t)$, 
\be
\frac{SNR_{\rm Ia}(t)}{({\rm 100 \, {\rm yr}})^{-1}} = 
A \left[ \frac{M_\star(t)}{10^{10} M_\odot} \right] + 
B \left[ \frac{\dot M_\star(t)}{10^{10} M_\odot \, {\rm Gyr}^{-1}} \right],
\ee
where $A$ and $B$ are dimensionless  constants that we fix with
observations (see also M05, eq.\ [2]).  The first of these terms is
dominant  in old stellar populations (and contains underluminous
1991bg-like SNe), while the second term is most important in
starbursts (and contains the brightest 1991T-like SNe).  To measure
$A,$ we use the recent observations from M05, who utilized detailed
$K$-band data to update the analysis presented by  Cappellaro \etal
(1999).  As there were  21 Type Ia SNe observed in E/S0
galaxies in this sample, and no instances of core-collapse supernovae,
we consider the SFR term to be negligible in this population.  This
gives $A = 4.4^{+1.6}_{-1.4} \times 10^{-2}$.

M05 showed that the  SN Ia rate was $0.35 \pm 0.08$ of the core
collapse rate in young stellar populations.  Since they arise from
massive, short-lived stars, the core-collapse  SN rate should directly
trace the SFR, and thus can be used to determine $B$. Presently, the
primary uncertainty in this measurement comes from relating the
core-collapse rate to the SFR.  We take two approaches to
determining $B$, fully aware that ongoing SN surveys will soon reduce
these uncertainties.  First we use the $z \leq 1.0$  core-collapse SN
rate density as measured by Dahl\' en \etal (2004), comparing it against
the SFR density as measured by  Giavalisco \etal
(2004), and considering only the statistical  error bars. This gives
$SNR_{cc}/\dot M_\star=(7.5 \pm 2.5) \times 10^{-3}$ $\msun^{-1}$ and
corresponds to a  $B$ value for SNe Ia  of $2.6 \pm 1.1,$  which we
adopt throughout this Letter.

An alternative approach is to use  the blue ($B-K \leq 2.6$)
population observed by M05, in which the measured SN Ia rate is
$0.86^{+0.45}_{-0.35}$ per 100 yr per $10^{10} \msun$ of stars.  The
colors of these starbursting galaxies are consistent with a 0.7 Gyr
population (M05), as computed from the population synthesis models of
Bruzual \& Charlot (2003).  Within the errors, this gives  $B =
1.2^{+0.7}_{-0.6},$ consistent with our first estimate.

\begin{figure}[tbp]
\centerline{\psfig{figure=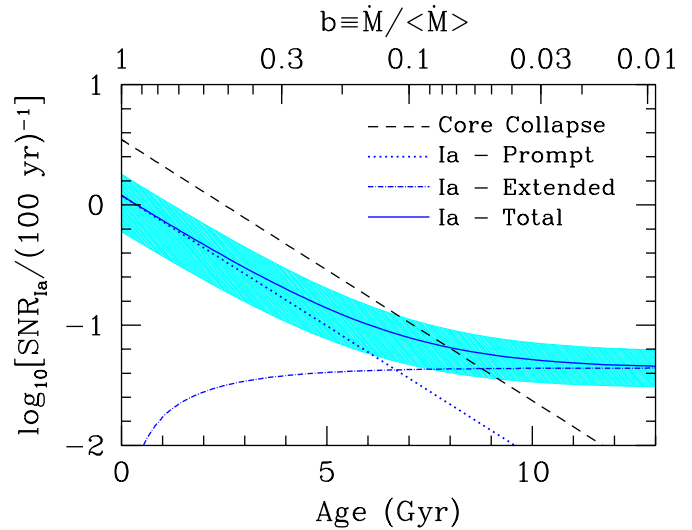,height=7.5cm}}
\caption{The supernova rate in a galaxy with a final stellar mass of
$10^{10}M_\odot$.  The solid line gives
our model predictions for  the Type-Ia SN rate (bracketed by 1 sigma errors), which is made up of the prompt
(dotted) and extended (dot-dashed) components.  The dashed line
gives the core-collapse SN rate.   In all cases we assume a star
formation rate  $\propto e^{-t/2{\rm Gyr}}.$ Choosing a different
characteristic star-formation decay time would rescale the time axis,
while leaving the Scalo $b$ values unchanged.}
\label{fig:mannuci}
\end{figure}

Our values of $A$ and $B$ directly yield the relative contributions of
these two components.   For example,  10 Gyrs after a starburst, only
20\% of all SNe Ia will have come from the extended piece.  Hence, in
our model, most of the SNe Ia over any galaxy's lifetime come from the
prompt contribution as originally suggested by Oemler \& Tinsley
(1979).

Our model also allows for a comparison between SN types, as
illustrated in Figure \ref{fig:mannuci}.   In our model, by
construction, the rate of core-collapse SNe  is approximately 3 times
the SN Ia rate in starbursting galaxies, while only SNe Ia are found
in galaxies without star formation.  For a galaxy with a total stellar
mass of $10^{10} \msun$ the transition between these two regimes
occurs at  $\dot M_\star \approx 0.1 \msun$ yr$^{-1},$ which corresponds
to a Scalo parameter $b \equiv \dot M_\star(t) / \left< \dot
M_\star(t) \right>=  \dot M_\star(t) t/M_\star(t) \approx 0.08$  or an
age of 7 Gyrs if we assume a star formation rate   $\propto
e^{-t/2{\rm Gyr}}$ as in M05.   

Finally, since $0.74 \msun$ of Fe is
expelled  in a typical SN Ia (Tsujimoto \etal 1995) while $0.062 \msun$
is expelled in a typical core-collapse SN (Hamuy 2003), in any given
starburst the overall SN Ia contribution to Fe production as compared with
core-collapse SNe in our model is approximately 3 to 1.
 
\section{Implications and Predictions}

We now apply this simple model to three important issues.  
In this section and below we adopt 
a Hubble constant of 70 km s$^{-1}$ Mpc$^{-1}$ and total matter 
and vacuum energy densities 
of $\Omega_m$ = 0.3 and $\Omega_\Lambda$ = 0.7
in units of the critical density (\eg Spergel \etal 2003). 
First we consider the intracluster medium (ICM) in galaxy clusters, which is
measured to have an  Fe content $\approx 0.3 Z_\odot$ (Baumgartner \etal
2003).  Although clusters are dominated by elliptical  galaxies,
models that combine the observed SN Ia rate in  ellipticals with the
total cluster stellar mass  result in Fe estimates that are roughly an
order of magnitude  too small (\eg Renzini \etal 1993; Renzini 2004).
While Maoz \& Gal-Yam (2004) were able to provide a single-component
resolution to this problem, they were not able to reconcile this
fit with SN Ia  measurements in field galaxies.
In fact, even ICM models that appeal to Fe production  by
pair-instability SNe from very massive primordial stars  (\eg
Lowenstein 2001) fall far short of the observed metallicity
(Scannapieco \etal 2003).

\begin{figure}[tbp]
\centerline{\psfig{figure=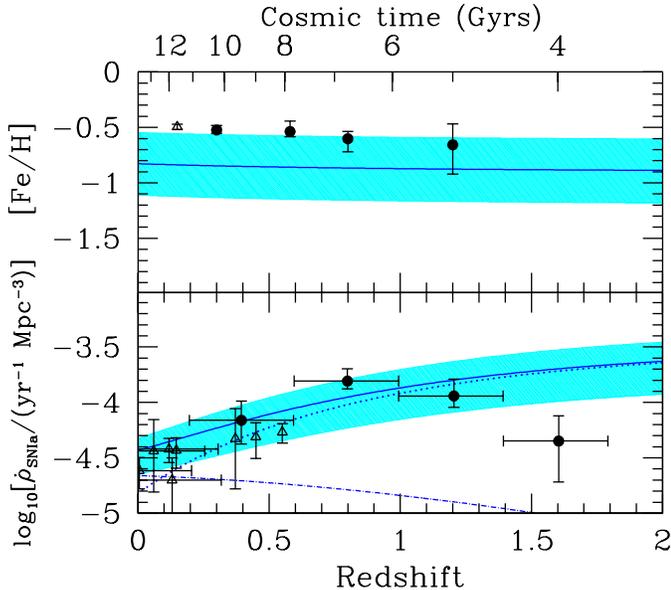,height=8.5cm}}
\caption{{\em Top:} [Fe/H] of the intracluster medium in $kT \geq
5$ keV galaxy clusters as a function of redshift. The solid line shows
the results of our two-component model, with the one-sigma errors
defining the  shaded region.  The low redshift point (open triangle)
is an average from the Baumgartner \etal (2003) sample, while the
higher redshift points (solid circles) are from Tozzi \etal (2003).
{\em Bottom:} Type Ia SNR density as a function of redshift (solid)
which is the sum of  the prompt (dotted) and extended (dot-dashed) 
components.   Again the solid
line corresponds to our two-component model,  with the one-sigma
errors given by the shaded region.  The lower redshift measurements
(open triangles) are taken (in order of increasing redshift) from
Cappellaro \etal (1999), Hardin \etal (2000), Blanc \etal (2004), Reiss
(2000), Pain \etal (1996), Tonry \etal \ (2003), and Pain \etal
(2002).  The higher redshift measurements (solid circles) are from
Dahl\' en \etal (2004).}
\label{fig:implications}
\end{figure}

Our two-component model, however, addresses this issue from a
different perspective.   The dominant
source of Fe is not the late-time SN Ia contribution, as 
observed in ellipticals, but rather the  prompt
contribution, which took place at high redshifts.  In the upper panel
of Figure \ref{fig:implications},  we plot the ICM metallicity as a
function of redshift,  assuming a star formation redshift of 3, and an
ICM to stellar mass ratio of  $M_{\rm ICM}/{M_{\star}} = 10 \pm 3,$ as
appropriate for 7 keV  clusters (Lin \etal 2003).  
Again, we take the core-collapse SN rate to directly trace the
SFR.  This results
in [Fe/H] values broadly consistent with observations and an order of
magnitude higher than previous estimates (Renzini 2004).  
Our model also naturally predicts the recently observed lack of [Fe/H]
evolution  with redshift (Tozzi \etal 2003),  a feature that does 
not appear in models dominated by late-time SNe Ia. 

Next we turn to the cosmic SN Ia rate density, or the number of Type
Ia SNe per year per comoving Mpc$^{3}$.  We adopt a cosmic star
formation rate density of  $\log_{10}[{\rm SFR} / (\msun \, {\rm
yr}^{-1} \, {\rm Mpc}^{-3})]   = -2.2 +3.9 \log_{10}(1+z) -3.0
[\log_{10}(1+z)]^2,$  which is a simple fit to the most recent
measurements  (Giavalisco \etal 2004; Bouwens \etal 2004).  The
resulting SN Ia rate density  is compared with observations
in the lower panel of Figure
\ref{fig:implications}, providing an excellent match,
except for the highest-redshift point from Dahl\' en \etal (2004).  This
is because at $z \approx 1$ our model is dominated by the prompt piece,
and no corresponding dip is seen in the SFR density at this redshift.
Furthermore,
requiring agreement with this point  is the source of the $2-4$ Gyr
delay-time derived by  Strolger \etal (2004).  Thus a strong {\em
prediction} of our model is that future observations will revise the
$z \approx 1.5$ measurement upward.  In fact, more detailed analyses of
the SN Ia rate density at $z > 1$ represent the single
best  way to falsify (or confirm) our approach.

\begin{figure}
\centerline{\psfig{figure=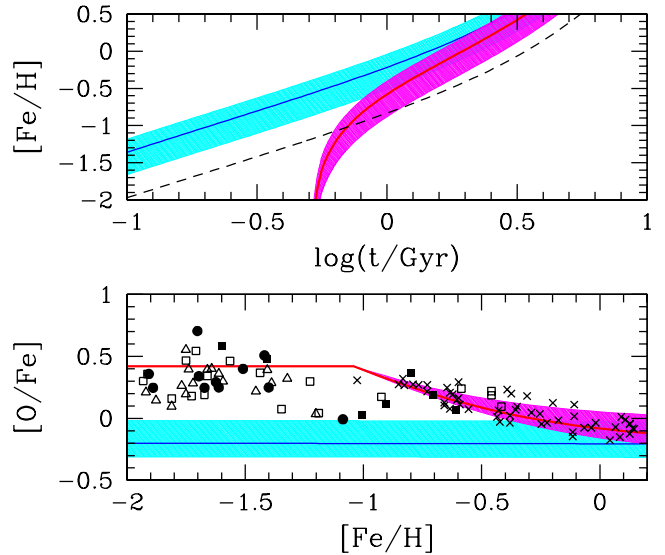,height=8.5cm}}
\caption{{\em Top:} Evolution of the Fe content in a closed-box
system as a function of the age of the stellar population, 
assuming an SFR $\propto e^{-t/2{\rm Gyr}}$.  Here the (thin)
upper solid line corresponds to our fiducial two-component model
(bracketed by the 1-sigma errors), while the dashed line gives the iron
provided by core-collapse SNe.  Finally, the (thick) lower sold line is the
result of our delayed two-component model.
{\em Bottom:} Variation of the [O/Fe] abundance ratio as a function of
metallicity, in a simple closed box model.  The (thin) lower line gives
the results of our fiducial model, while the upper (thick) line corresponds
to the two-component model with a $0.7$ Gyr delay in the prompt 
component.  The data points are taken
from observations of Galactic disk and halo stars compiled in
McWilliam (1997),  following the symbol convention used in his Figure 3.}
\label{fig:earlyevolution}
\end{figure}

Finally we turn to the measured abundance ratios of Galactic stars,
which  probe the relationship between Type Ia and core-collapse SNe at
short times.  In particular we compare iron with oxygen, an
alpha-element that is synthesized primarily  in core-collapse SNe.  We
take $0.14 \msun$ of oxygen per SN Ia (Tsujimoto \etal 1995),  $1.2
\msun$ oxygen per core-collapse SN (a Salpeter inital mass function
average  computed in Scannapieco \etal 2003), and a closed-box model.
Following M05 we adopt  a star formation rate  $\dot M_\star = M_{\rm
gas} \exp[-t/2 {\rm Gyr}]/ 2 {\rm Gyr},$ which results in the [Fe/H]
and [O/Fe] values shown in Figure \ref{fig:earlyevolution}.

As the overall Ia Fe contribution as compared with core-collapse SNe
in our model is nearly 3 to 1, [O/Fe] should drop by  a factor of 3,
as observed.    The value of [Fe/H] where this drop occurs, however,
depends on the star formation model, and requires a short delay in the
prompt component. Assuming that both the core-collapse and the prompt
Type Ia contributions exactly trace the SFR would fix the [O/Fe]
values to a constant, in conflict with measurements of metal-poor halo
stars. On the other hand, delaying the prompt component by $0.7$ Gyr
allows core-collapse SNe to briefly dominate the initial gas
enrichment, but has no impact on the SN Ia distribution on the timescales
probed by other measurements.   The timescale of this delay is
proportional to the assumed SFR decay time, and a model with  an SFR
decay time of $1.0$ Gyr and a prompt SN Ia delay of $0.35$ Gyr  would
give equivalent  [O/Fe] values. In any case, this delay is so short on
cosmic times that we will continue to refer to this component as
``prompt.''

\section{Comparison with Other Work}

We now compare our approach with other single-component fits.  In
particular we consider three possible SN Ia delay functions: an
exponential model  in which $\Phi(\Delta t) = C \exp(-\Delta
t/\tau)/\tau$ (Tutukov \& Yungelson 1994; Madau \etal 1998; Gal-Yam \&
Maoz 2004), and two Gaussian models in which $\Phi(\Delta t) =  C (2
\pi \sigma^2)^{-1/2}  \exp[(\Delta t -\tau)^2/(2 \sigma^2)]$ and
$\sigma$ is either ``narrow'' ($\sigma = 0.2 \tau)$ or ``wide''
($\sigma = 0.5 \tau$) (Dahl\' en \& Fransson 1999;  Strolger \etal 2004).
Each of these functions has two free  parameters: a delay time $\tau$
and a normalization $C$; which we fit to the number of SNe Ia in the
youngest ($B-K < 2.6$) and most evolved ($B-K > 4.1$) galaxies
measured by M05.  As in that study, we model both galaxy types 
with  $\dot M_\star \propto \exp(-t/2 {\rm Gyr})$ with an age
of $0.75 \pm 0.25$ Gyrs in the  $B-K < 2.6$ population and $10.5 \pm
1.5$ Gyrs in the $B-K > 4.1$ population, as is consistent with their
observed colors  and core-collapse SNe rates.  Note that these SFRs
are averages over entire populations, 
rather than histories of individual galaxies.   
Within the one-$\sigma$
errors, we find $\tau$ values of $0.5-1.6$,  $0.6-1.0$, and
$0.5-1.3$ Gyr, for the exponential, narrow Gaussian, and wide Gaussian
models, respectively.

Applying these fits to the full range of galaxy populations  measured
by M05 indicates that  the proper number of SNe Ia in old galaxies and
starbursting galaxies  is obtained only at the expense of a large
number of SNe Ia  in galaxies of intermediate age.  For example in a
$t=5$ Gyr population, all three models predict $\geq$ 2 SNe Ia per 100
yr per $10^{10} \msun$, while the measured value is
$0.19^{+0.08}_{-0.07}$ (M05).  Furthermore the ratio of Type Ia to
core-collapse SNe at 5 Gyr is $\geq 4$, while the measured ratios are
$\leq 1/3.$ On the other hand, our two-component model, shown as the
solid line in Figure \ref{fig:mannuci},  falls within the range of
observed values at all ages.

\vspace{.1in}

\section{Conclusions}

Our two-component model is motivated by the observed dichotomy between
the environments of the brightest (1991T-like)  and the faintest
(1991bg-like) SNe Ia.   Yet in some sense it is simply an application
of the  delay function formalism,  in which the SN Ia rate is described
as a convolution of an unknown function  with the overall star
formation history.  Other models  were limited by the assumption of a
single ``delay time''  and had difficulties in reconciling the Fe
content in clusters with the ratio of core-collapse to Type Ia SNe as a
function of galaxy age (\eg Maoz \& Gal-Yam 2004). Our model solves
this problem because it is dominated by a prompt component, but allows
significant numbers of SNe Ia to occur  at late times. In addition, it
produces the observed cosmic SN Ia rate to $z \leq 1$, while also
fitting observations of E/S0s.  The strongest test of this model is
the measurement of the SN Ia rate at $z > 1$, which we predict to be in
the range $1-3.5 \times 10^{-4}$ yr$^{-1}$ Mpc$^{-3}$ at $z=2$.

Of course, star formation does occur in some elliptical galaxies, and
as shown in Figure 1, our simple model predicts that prompt SNe Ia will be 
the dominant component in such objects if  $\dot M_\star/ \left< \dot
M_\star \right> \gtrsim 0.1$.  An example of such a case  is the
slow-declining Type Ia SN 1998es,  which occurred in the early-type
galaxy  NGC632.  While this galaxy is fairly red (B-K $>$ 3), spectral
observations uncover significant star formation (Gallagher \etal
2005).   Conversely, underluminous SNe Ia should occasionally be found
in star forming galaxies, such as the rapidly-declining SN 1999by.
While the host galaxy of this SN Ia is an Sb galaxy, imaging shows
that it took place in the old population of halo stars (Gallagher
\etal 2005).

SNe Ia play a pivotal role in astrophysics, and thus our two-component model
has many implications.  It allows for an updated assessment of Clayton \&
Silk's (1969) hypothesis that the $\gamma$-rays from radioactive
decays explain the extragalactic MeV background (see Watanabe \etal 1999; 
Ruiz-Lapuente \etal 2001;  Ahn \etal 2005). It highlights
the usefulness of measurements that constrain Type Ia evolution at
short time scales, such as studies of the distribution
of SNe Ia relative to spiral arms (Maza \& van den Bergh 1976;
Bartunov \etal 1994; McMillan \& Ciardullo 1996; Petrosian \etal 2005) and
the Fe content of high-redshift quasars (Barth \etal 2003; Dietrich
\etal 2003). It stresses the importance of early SNe Ia in ICM
enrichment and exposes the limitations of one-component fits.

More generally, our model implies that over a Hubble time, $\approx
80\%$ of the SNe Ia  from any  galaxy will occur within a Gyr of the
initial starburst.  The remaining $20\%$ occur in a delayed fashion,
clearly extending to times $\gsim 10$ Gyrs. This alludes to multiple
progenitor scenarios: one that occurs ``promptly,'' within a Gyr, and
another that can occur a Hubble time after star formation.  Perhaps
this is  not surprising given the openly-debated range of
possibilities  (\eg Branch et al. 1995;  Greggio 2005) and the
evidence for dominance (or absence) of some extreme SNe Ia in certain
galaxy types.  Deeper physical insights into how the age or
metallicity of the accreting white dwarf might naturally cause this
large range of diversity awaits theoretical work.

\acknowledgments 

We thank Avishay Gal-Yam and the anonymous referee for comments.
This work was supported by the NSF under  
grants PHY99-07949 and AST02-05956. 

\fontsize{10}{10pt}\selectfont


\end{document}